\documentclass[reprint, amsmath,amssymb,aps,prb,floats,floatfix]{revtex4-2}

\usepackage{graphicx}
\usepackage{bm}
\usepackage{bbm}
\usepackage{xcolor}

\begin{document}

\title{Fock-space geometry and strong correlations in many-body localized systems}

\author{Christian P. Chen}
\affiliation{Department of Physics, Lancaster University, Lancaster, LA1 4YB, United Kingdom}
\author{Henning Schomerus}
\affiliation{Department of Physics, Lancaster University, Lancaster, LA1 4YB, United Kingdom}

\date{\today}
\begin{abstract}
We adopt a geometric perspective on Fock space to provide two complementary insights into the eigenstates in many-body-localized fermionic systems. On the one hand, individual  many-body-localized eigenstates are well approximated by a Slater determinant of single-particle orbitals. On the other hand, the orbitals of different eigenstates in a given system display a varying, and generally imperfect, degree of compatibility, as we quantify by a measure based on the
projectors onto the corresponding single-particle subspaces.
We study this incompatibility between states of fixed and differing particle number, as well as  inside and outside the many-body-localized regime.
This gives detailed insights into the emergence and strongly correlated nature of quasiparticle-like excitations
in many-body localized systems, revealing intricate correlations between states of different particle number down to the level of individual realizations.

\end{abstract}

\maketitle

\section{Introduction}

Many-body localized (MBL) systems constitute a broad class of closed quantum systems that fail to equilibrate under their own dynamics \cite{annurev-conmatphys-031214-014726,annurev-conmatphys-031214-014701,RevModPhys.91.021001,PhysRevB.82.174411,PhysRevB.75.155111,PhysRevB.77.064426}. This behaviour manifests itself as a violation of the eigenstate thermalization hypothesis \cite{PhysRevA.43.2046,PhysRevE.50.888}, a hypothesis that more conventional, ergodic systems obey.
A pressing question \cite{ABANIN2021168415} is whether the MBL phase can be explored equally via features of the wavefunctions, such as in the area law of entanglement \cite{PhysRevB.75.155111,PhysRevB.82.174411}, and via energy-level statistics \cite{PhysRevE.102.062144}, as well as how firmly these approaches tie into key theoretical concepts.
In particular, wavefunction-based characterizations so far approach each eigenstate individually, while energy-level statistics are generally based on energy-level spacings, and thereby involve the interplay of several states \cite{PhysRevB.75.155111,PhysRevB.93.041424,PhysRevB.94.144201}. This leaves a conceptual gap between the most reliable characterizations of MBL and the most influential theoretical framework, the concept of robust emergent integrability \cite{PhysRevLett.111.127201,PhysRevB.90.174202,j.nuclphysb.2014.12.014,s10955-016-1508-x}, which ties  the MBL phase transition to the emergence of an extensive set of local conserved quantities that are commonly referred to as l-bits \cite{PhysRevB.90.174202,PhysRevLett.111.127201,j.nuclphysb.2014.12.014,s10955-016-1508-x,PhysRevLett.117.027201,andp.201600278,andp.201600322}. These l-bits can be seen as highly structured constraints on the complete set of eigenstates of any given system, both energetically as well as in terms of their wavefunctions.
In principle, an all-encompassing description of MBL should therefore also allow to develop expectations for the \emph{joint characteristics of the wavefunctions} of a given system.

In this work, we set out a framework to formulate and verify such expectations for the paradigmatic case of fermionic systems, and use this to uncover their strongly correlated nature from a transparent geometric perspective.
Our framework rests on the observation that for these systems, the MBL wavefunctions are closely approximated by Slater determinants, a realization that previously proved useful to lift Fock-space localization \cite{PhysRevB.101.134202,roy2021fock} into real space by means of the one-particle density matrix \cite{PhysRevLett.115.046603,andp.201600356,PhysRevB.97.104406,SciPostPhys.4.1.002,SciPostPhys.4.6.038,SciPostPhys.6.4.050,PhysRevResearch.2.023118,orito2020effects,PhysRevA.101.063617,PhysRevB.96.060202,hopjan2020scaling}.
Here, we propose to capture this structure by considering the span of single-particle orbitals whose Slater determinant provides the largest overlap to a given eigenstate---known in quantum chemistry as Brueckner orbitals \cite{PhysRev.97.1344,PhysRev.103.1008,PhysRev.109.1632}---and show that this indeed provides a natural avenue to study the unexplored geometric interrelations between the wavefunctions of a given system.

Our key insights are the following: While individual MBL wavefunctions are indeed  well approximated by a single Slater determinant, when comparing different eigenstates \emph{of the same system} the underlying orbitals are generally mutually incompatible with each other.
We  define and quantify this mutual incompatibility precisely using a natural geometric measure, based on the
projection operators of the occupied Brueckner orbitals from different eigenstates. We then establish the finite extent of this incompatibility in a model system, use random-matrix theory as a benchmark to show that it scales systematically with system size, and develop this picture further by varying the disorder and interaction strength.
Finally, we relate the observed incompatibility to features of the aforementioned l-bit operators, whose systematical dressing is revealed by geometric correlations of states with different particle numbers.
The proposed framework therefore provides a natural perspective on the emergence of MBL that complements traditional insights, revealing strongly correlated features that can be precisely quantified.

\section{Brueckner orbitals and orbital incompatibility}
In a sector of $N$ fermions,
Brueckner orbitals are the name given to a set $S_m=\{\chi_n^{(m)}\}_{n=1}^N$ of single-particle orbitals which maximize the overlap \begin{equation}
	\label{eqn:Overlap_1}
	\mathcal{I}_{m}\equiv|\langle\psi_m|S_m\rangle|^2
\end{equation}
between the Slater determinant  $|S_m\rangle$  constructed from these orbitals  and a given many-body eigenstate $|\psi_m\rangle$ of the system.
These orbitals enjoy a wide range of applications in quantum-chemical approaches to interacting quantum systems \cite{PhysRev.97.1344,PhysRev.103.1008,PhysRev.109.1632,PhysRev.178.137}, and their use can also be extended to a systematic approximation by a series of Slater determinants \cite{PhysRevA.89.012504,PhysRevA.94.032513}.
For this work, an important feature of these orbitals is the fact that the Slater determinant $|S_m\rangle$ only depends on the subspace spanned by the Brueckner orbitals, so all their relevant features are uniquely captured by the projector $P_m$ onto this subspace.
In general, the Brueckner orbitals are distinct from the well-known natural orbitals $\varphi_\alpha^{(m)}$, which in a discrete basis of single-particle states with fermionic annihilation operators $c_i$  are obtained by diagonalizing the one-particle density matrix (OPDM)
	$\rho_{ij}^{(m)} \equiv \langle\psi_{m}|c_{i}^{\dag}c_{j}|\psi_{m}\rangle$,
where the eigenvalues $n_{\alpha}^{(m)}$ represent the  occupations of these natural orbitals.
When the overlap $\mathcal{I}_{m}$ is moderate or small, both types of orbitals can be very different to each other, to the extent that constructing the Brueckner orbitals from the natural orbitals can be a numerically unfavourable task \cite{PhysRev.178.137,PhysRevA.89.012504,PhysRevA.94.032513}. However, the situation changes when the overlap  $\mathcal{I}_{m}$ is large, and for $\mathcal{I}_{m}\to 1$,
$\rho^{(m)} \to P_m$.
Therefore, we can write these projectors explicitly as $(P_{m})_{ij} = \langle S_{m}|c_{i}^{\dag}c_{j}|S_{m}\rangle$.
In turn, from the perspective of the OPDM this limit is reached when the occupations  $n_{\alpha}^{(m)}\to 0,1$.
This indeed applies well to individual eigenstates inside the many-body-localized regime \cite{PhysRevLett.115.046603,andp.201600356,PhysRevB.97.104406}, and we will expand on this observation further below.

Turning our attention to the relation between different eigenstates, our key premise is the following
observation: \emph{Even when two (or perhaps all) eigenstates of an interacting system are exact Slater determinants, this does not guarantee that there is a common single-particle basis from which one can form the Brueckner orbitals of these  eigenstates}.
In  particular, such a single-particle basis is not guaranteed by the mere requirement that the corresponding many-body Slater determinants are orthogonal to each other.
This is in contrast to noninteracting systems, where the single-particle eigenstates provide a joint set from which one can choose the Brueckner orbitals.
Indeed, it is easy to quantify and detect this \emph{orbital incompatibility}, as it can be established already on the level of pairs of many-body eigenstates $|\psi_{l}\rangle$ and  $|\psi_{m}\rangle$. When the orbitals of both states are compatible with each other, their projectors $P_l$ and $P_m$ commute. We therefore introduce the following incompatibility measure,
\begin{equation}
\label{eqn:OPDM_5}
\mathcal{C}_{ml} \equiv
\sqrt{\mathrm{tr}\,[P_l,P_m][P_m,P_l]}= \sqrt{2\mathrm{tr}\,(P_lP_m-(P_lP_m)^2)},
\end{equation}
which mathematically corresponds to the Frobenius norm of the commutator of both projectors.
As the projectors $P_m$ characterize $N$-dimensional subspaces in the single-particle Hilbert space $H_1$---mathematically, the Grassmannian  $G(H_1,N)$---this provides us with a simple geometric picture of the interrelation between different wavefunctions of a given  MBL system.

\begin{figure}[t!]
	\includegraphics[width=\linewidth]{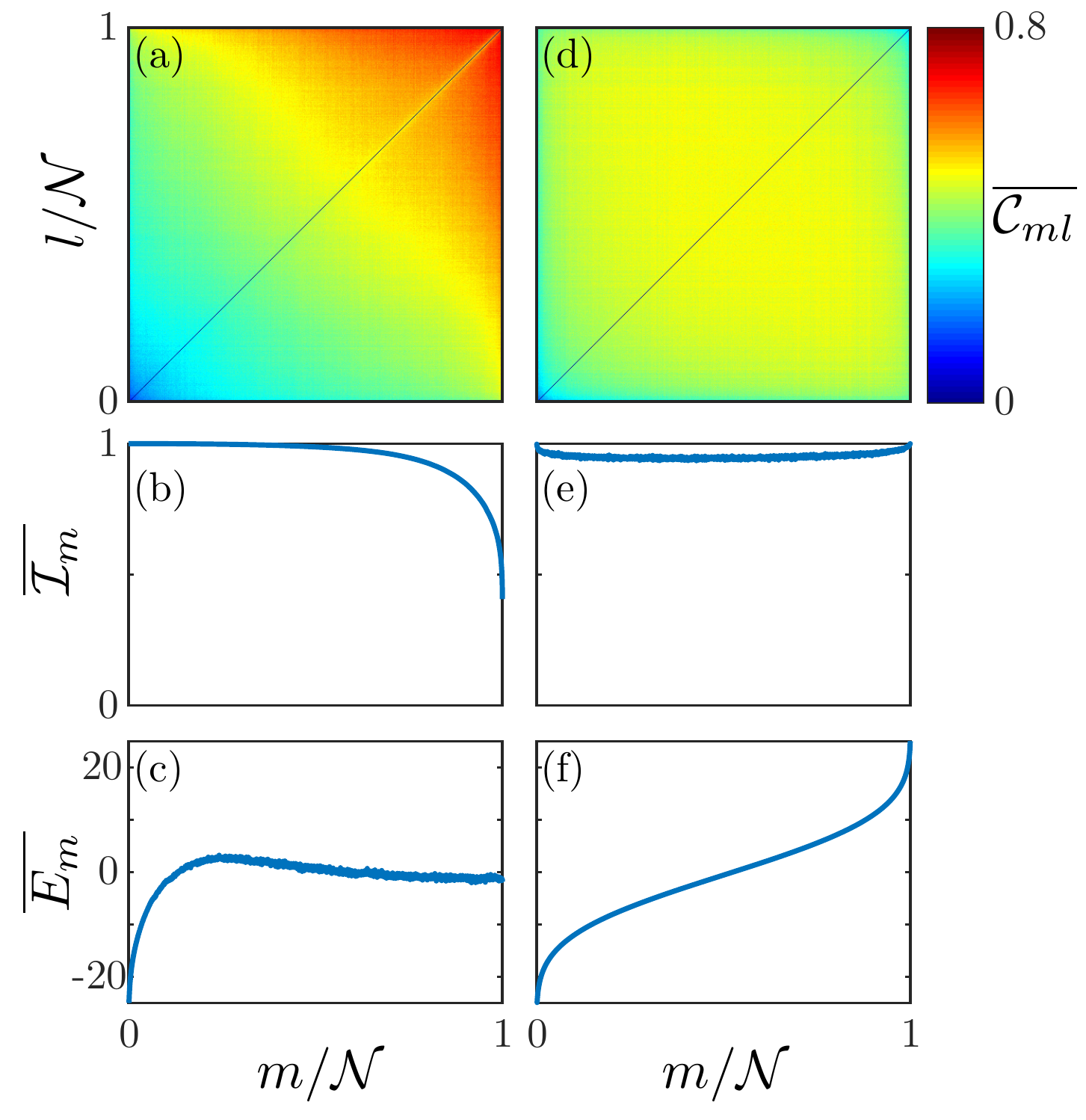}
	\caption{
Geometric features of eigenstates in the many-body localized system \eqref{eqn:Htot_1} (half-filled system of size $L=14$, with $V=1.5$, $W=8$; total number of states denoted as $\mathcal{N}$). In the left column, states are ordered by descending overlap $\mathcal{I}_m$ with a single Slater determinant, while in the right column they are ordered by ascending energy $E_m$; see panels (b,c,f,g) for the averaged behaviour of these quantities. The observed large overlaps $\mathcal{I}_m$ show that a large fraction of states are well approximated by a single Slater determinant.
Panels (a,d) show the
averaged incompatibility $\mathcal{C}_{ml}$ defined in Eq. \eqref{eqn:OPDM_5}, which quantifies the geometric interrelation of the states. This incompatibility is appreciable even for states with $\mathcal{I}_m\approx 1$, which are very precisely approximated by a single Slater determinant.
}
	\label{fig:OCM_L14_N7}
\end{figure}

\section{Application}
To  illustrate the geometric framework described above in a specific setting, we place it into the context of a paradigmatic fermionic model exhibiting many-body localization.
This is a system consisting of $N$ locally interacting spinless fermions confined to a one-dimensional lattice with $L$ sites and subjected to a disordered onsite potential, where the Hamiltonian is written as
\begin{align}
	\label{eqn:Htot_1}
	H = & -\frac{1}{2}\sum_{i=1}^{L}\left(c_{i}^{\dag}c_{i+1}+\mathrm{h.c.}\right)
	+\sum_{i=1}^{L}\varepsilon_{i}\left( n_{i}-\frac{1}{2}\right) \nonumber\\
	& +V\sum_{i=1}^{L}\left( n_{i}-\frac{1}{2}\right)\left( n_{i+1}-\frac{1}{2}\right).
\end{align}
Here $c_{i}$ destroys a fermion on site $i$, and $n_{i} = c_{i}^{\dag}c_{i}$ is the associated number operator, while we set
$c_{i+L}\equiv c_i$ to enforce periodic boundary conditions. Disorder is introduced via a uniformly distributed uncorrelated onsite potential $\varepsilon_{i}\in [-W,W]$, where $W$ denotes the disorder strength, while the interactions of strength $V$ depend on the occupations of neighbouring sites.
Earlier investigations of this system and its equivalent formulations in terms of interacting spins
 have established that for moderate interactions ($V\sim 1$) and system sizes the
 critical disorder
strength $W_c$ of the MBL phase transition takes values in the range between 3 and 4  \cite{Santos_2004,PhysRevB.69.132404,PhysRevB.75.155111,PhysRevB.77.064426,PhysRevB.82.174411,De_Luca_2013,PhysRevB.91.081103,PhysRevLett.114.100601,PhysRevLett.115.046603},
even though recent works indicate that in the thermodynamic limit of very large systems, the transition occurs at larger disorder strengths \cite{morningstar2021avalanches,sels2021markovian}.

\begin{figure}[t!]
	\includegraphics[width=\linewidth]{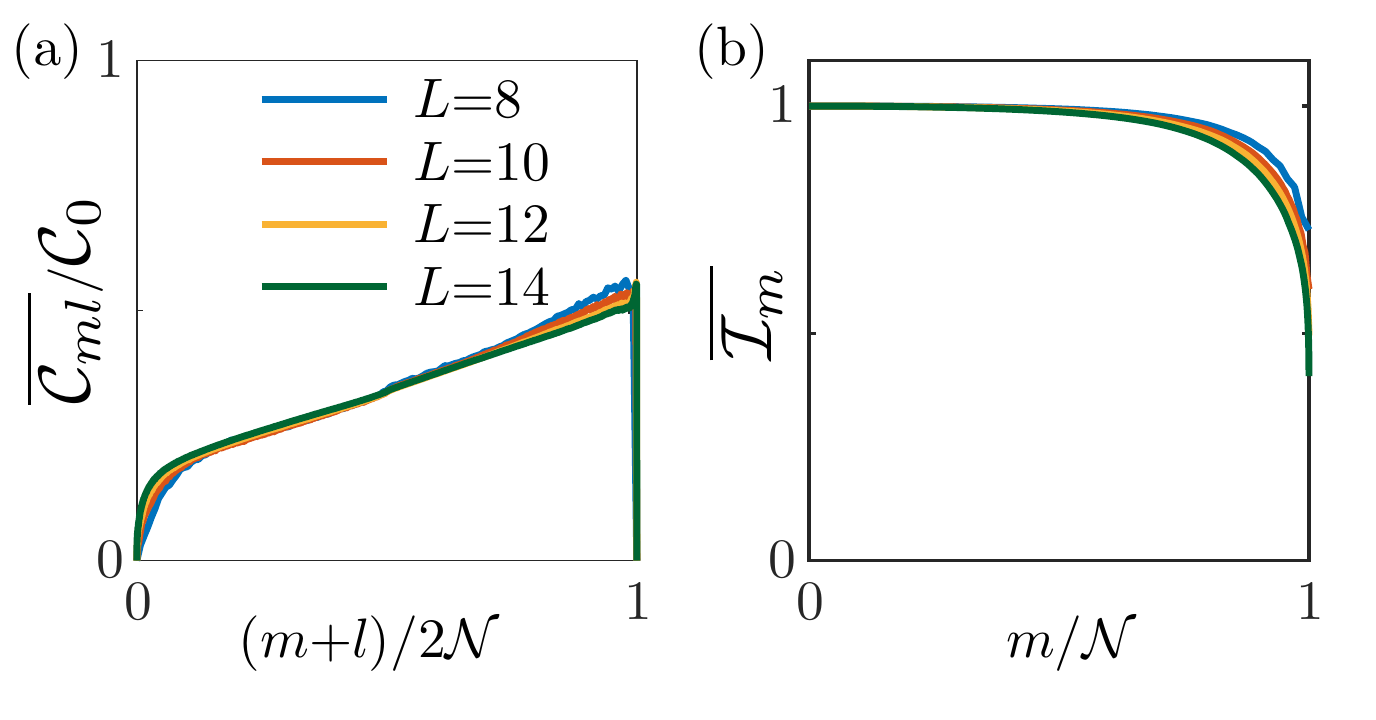}
	\caption{Panel (a) shows the orbital incompatibility $\mathcal{C}_{ml}$ for half-filled systems as in Fig.~\ref{fig:OCM_L14_N7}, but for
different sizes  $L = 8,10,12,14$ and additionally
averaged over pairs with fixed $l+m$ (states are  ordered by descending overlap $\mathcal{I}_m$).
The collapse of data in units of the random-matrix benchmark $\mathcal{C}_0=\sqrt{L/8}$
reveals that this incompatibility scales systematically as in a random system, and takes appreciable values. Panel (b) shows that the overlap $\mathcal{I}_m$ itself only weakly depends on the system size.}
	\label{fig:SC_L14_N7}
\end{figure}

To establish the incompatibility of MBL states numerically, we first calculate the Brueckner orbitals for each many-body eigenstate in a given realization
\footnote{Brueckner orbitals can be constructed systematically in several ways. Here, we adopt the method of  Ref.~\cite{PhysRevA.89.012504}, where one starts from a set of trial orbitals and updates these iteratively one at a time, by which $\mathcal{I}_{m}$ increases  monotonously. This method can also be extended to approximate the state using $M>N$ orbitals entering several Slater determinants.}, and determine from these  the overlaps $\mathcal{I}_m$. We then order the states either by overlap $\mathcal{I}_m$ or energy $E_m$, and next perform the disorder averages of these quantities, as well as of the incompatibility  measures $\mathcal{C}_{ml}$ for all pairs of eigenstates. All averages are carried out over $10^{3}$ realizations.

The results of this procedure in a many-body localized system ($V=1.5$, $W=8$) of length $L=14$ at half filling ($N=7$) are shown in Fig.~\ref{fig:OCM_L14_N7}.
In panels (a-c) the states are ordered by overlap.
The averaged overlaps in panel (b) confirm that,
as anticipated, a large fraction of states are exceedingly well approximated by a Slater determinant.  Panel (a) then verifies that nonetheless, these states generically have noticeably incompatible orbitals, with the incompatibility roughly constant along the antidiagonal. The incompatibility  is less pronounced for the states with the very largest overlaps, which as seen in panel (c) arise from the edge of the energy spectrum. Panels (d) and (e) show that these features are otherwise roughly independent of energy, provided again that we avoid the edges of the energy spectrum, according to which the states are ordered there [see panel (f)].

As shown in Fig.~\ref{fig:SC_L14_N7}, the observed incompatibility is indeed sizable and scales systematically with the system size. As a benchmark, we consider the extreme case of a pair of states with randomly incompatible orbitals, and hence mutually uncorrelated projectors. Applying random-matrix theory (RMT; see Appendix \ref{app:rmt}),
we find that for such a pair of states, the squared incompatibility averages to $\overline{\mathcal{C}_{ml}^2}|_{\mathrm{RMT}}=2N^2(L-N)^2/(L-1)L(L+\eta)$, where $\eta=1$ for complex orbitals and $\eta=2$ for real orbitals. Hence, as self-averaging suppresses statistical fluctuations, for a  half-filled large system $\overline{\mathcal{C}_{ml}}|_{\mathrm{RMT}}\sim \sqrt{L/8}\equiv \mathcal{C}_{0}$.
To compare the MBL system with this benchmark, we take the result from Fig.~\ref{fig:OCM_L14_N7}(a) averaged over the antidiagonals, and supplement this with the corresponding results for half-filled systems of size $L=8,10,12$.
Figure~\ref{fig:SC_L14_N7}(a) then confirms that the incompatibility in the MBL system is indeed sizeable,
and in particular scales reliably as $\overline{\mathcal{C}_{ml}}=O(L^{1/2})$, with a slight tendency to flatten out for larger systems.

\begin{figure}[t!]
	\includegraphics[width=\linewidth]{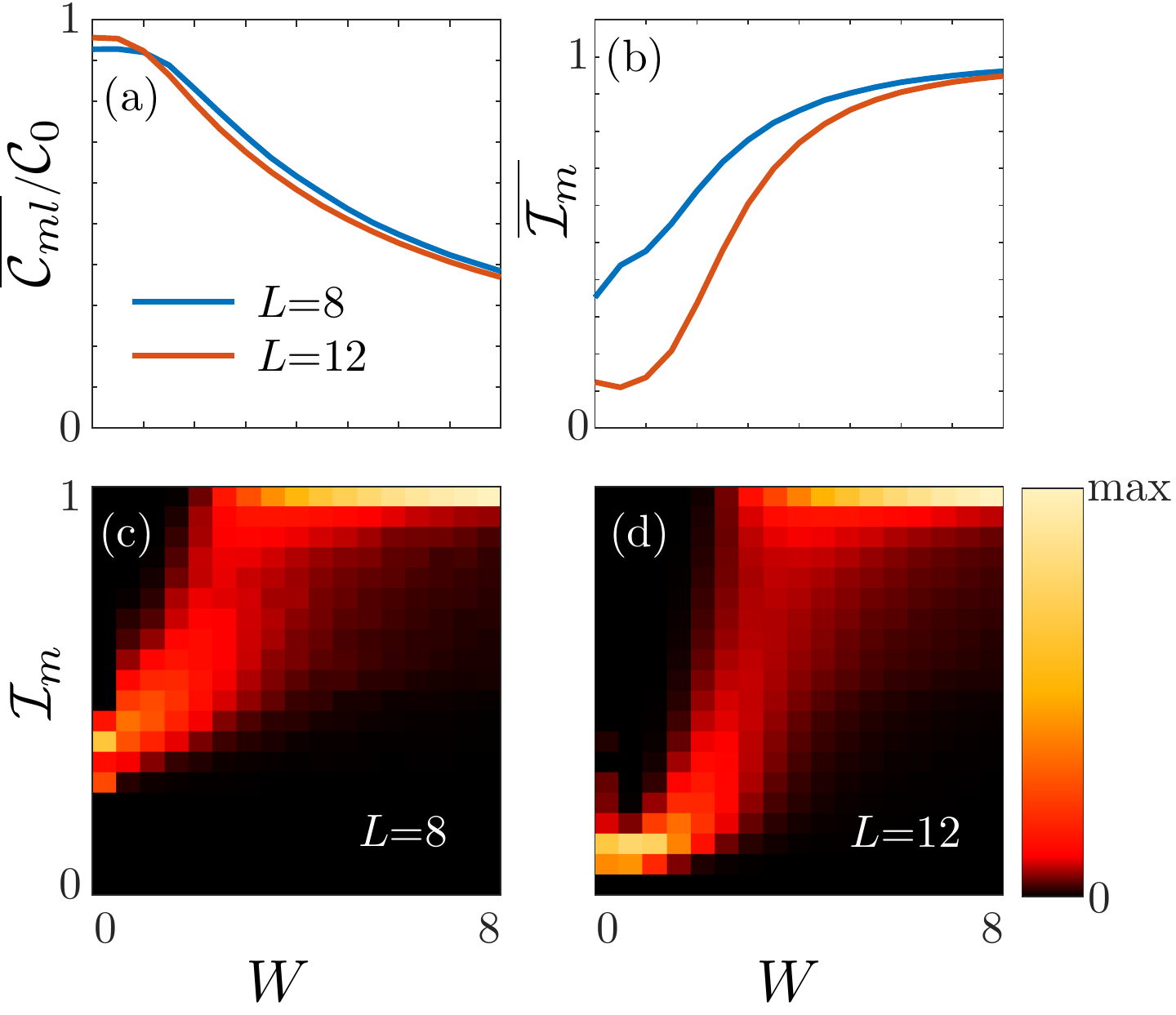}
	\caption{Dependence of (a) the averaged incompatibility $\mathcal{C}_{ml}$ and (b) the overlap $\mathcal{I}_m$ on the disorder strength $W$ (half-filled systems of size $L=8,12$ at fixed $V=1.5$, obtained from eigenstates from the middle $10\%$ of the energy spectrum). Panels (c,d) display the color-coded probability density function of $\mathcal{I}_m$ for both system sizes.}
	\label{fig:CompTransition_1}
\end{figure}

Figure \ref{fig:CompTransition_1} illuminates how these characteristic features  develop as a function of the disorder strength, where we focus on states from the central 10\% of the energy spectrum. At values of $W\gtrsim 4$, where many-body localization sets in, a noticeable fraction of these states become very well approximated by a single Slater determinant. This occurs roughly independently of the system size, as can be seen both from the averages shown in panel (b) as well as, in particular, in the colour-coded probability-density functions (pdfs) $P(\mathcal{I}_m)$ in panels (c,d). For smaller values of $W$, hence, in the ergodic phase, such well-approximated states are rare, and the pdf is strongly size dependent. On the other hand, the incompatibility measure $\mathcal{C}_{ml}$ retains its systematic scaling for all disorder strengths, approaching its RMT limiting value $\mathcal{C}_{0}$ for very weak disorder.
Most importantly, this measure is again still sizeable deep in the MBL phase, dropping only gradually to $\approx \mathcal{C}_{0}/2$  at $W=8$.

\begin{figure}[t!]
	\includegraphics[width=\linewidth]{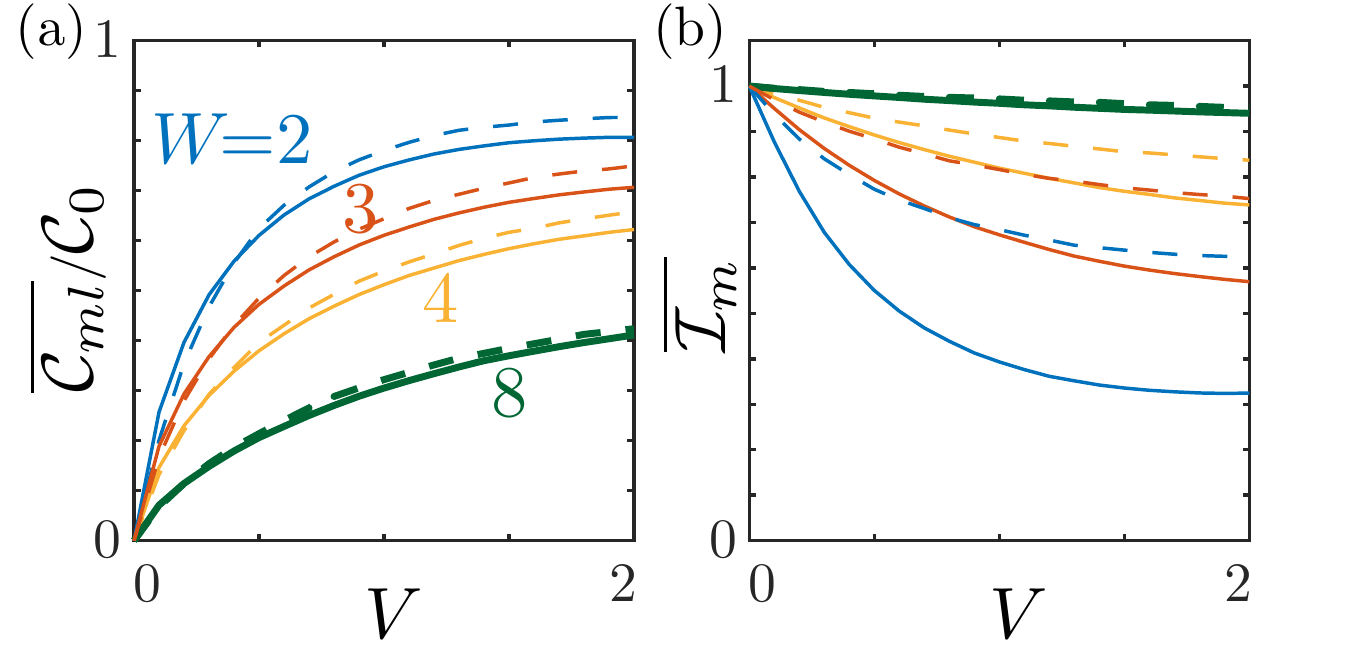}
	\caption{Dependence of (a) the averaged incompatibility $\mathcal{C}_{ml}$ and (b) the overlap $\mathcal{I}_m$ on the interaction strength $V$ (half-filled systems of size $L=12$, solid lines, and $L=8$, dashed lines, obtained from eigenstates from the middle $10\%$ of the energy spectrum).
The thick lines show data for $W=8$, where the system crosses over from Anderson localization to many-body localization. The thin lines contrast this to cases where the system transitions towards ergodic behaviour.}
	\label{fig:CompAnd_1}
\end{figure}

Figure \ref{fig:CompAnd_1} further explores how the incompatibility develops as interactions $V$ are switched on when one moves from Anderson localization to many-body localization ($W=8$, thick curve), and contrasts this with systems developing ergodic many-body dynamics  ($W=2,3,4$, thin curves).
In the ergodic systems, the  overlaps $\mathcal{I}_m$ drop significantly with increasing interactions, and depend noticeably on system size.
In contrast, in the cross-over to the MBL regime for $W=8$, the  overlaps $\mathcal{I}_m$ remain high while the incompatibility $\mathcal{C}_{ml}$ increases, with both quantities being only weakly dependent on the system size.

\section{Connection to l-bits}
\subsection{Numerical evidence for quasiparticle dressing}
So far, we have used the Brueckner orbitals to provide a direct geometric view of the relation between different many-body localized eigenstates.
In the remainder, we highlight the utility of this framework to further illuminate the general phenomenology of these systems, and formulate new questions.
This applies particularly to the notion of emergent integrability involving an extensive set of local conserved quantities, the aforementioned l-bits.
In the present work, these l-bits provide useful context when one interprets them as quasiparticle densities, $\mathcal{Q}_{i}=q_{i}^{\dag}q_{i}$, and seeks to construct the quasiparticle operators \cite{andp.201600356}
\begin{equation}
	\label{eqn:Quasi_2}
	q_{i}^{\dag} = p_{i}^{\dag} + \sum_{l,m,n}K_{lmn}p_{l}^{\dag}p_{m}^{\dag}p_{n} + \ldots
\end{equation}
in terms of single-particle operators $p_i$.
In the noninteracting case, where $q_{i}^{\dag} = p_{i}^{\dag}$, all eigenstates are Slater determinants constructed from a shared set of single-particle orbitals.
In a many-body localized system, higher-order terms must be taken into account, and a key guiding question is then whether this expansion is perturbative in nature.
Focusing on the first term in the series and assuming that it dominates, this would require the eigenstates of the system to be well approximated by Slater determinants. But as we have seen, this is not a sufficient condition: even though the eigenstates are all well approximated by Slater determinants, these Slater determinants cannot be constructed from a common set of single-particle orbitals, revealing that the construction of the l-bits is of a nonperturbative nature.

To further quantify this,
we consider the relation of eigenstates $|\psi_m\rangle$ with $N=L/2$ particles  to eigenstates $|\psi_l\rangle$ with $N=L/2-1$  particles.
Some of these eigenstates should be related by the application of a single quasiparticle operator, $|\psi_m\rangle= q_{i}^{\dag}|\psi_l\rangle$. In terms of the projectors $P_m$ and $P_l$,
the compatibility of these states will then depend on how much the higher-order dressing terms in $q_{i}^{\dag}$ fall into the space of orbitals that are already occupied in $|\psi_l\rangle$. If they fully do, the two states will have the same overlap with a single Slater determinant, $\mathcal{I}_m=\mathcal{I}_l$, and be highly compatible.

\begin{figure}[t!]
	\includegraphics[width=\linewidth]{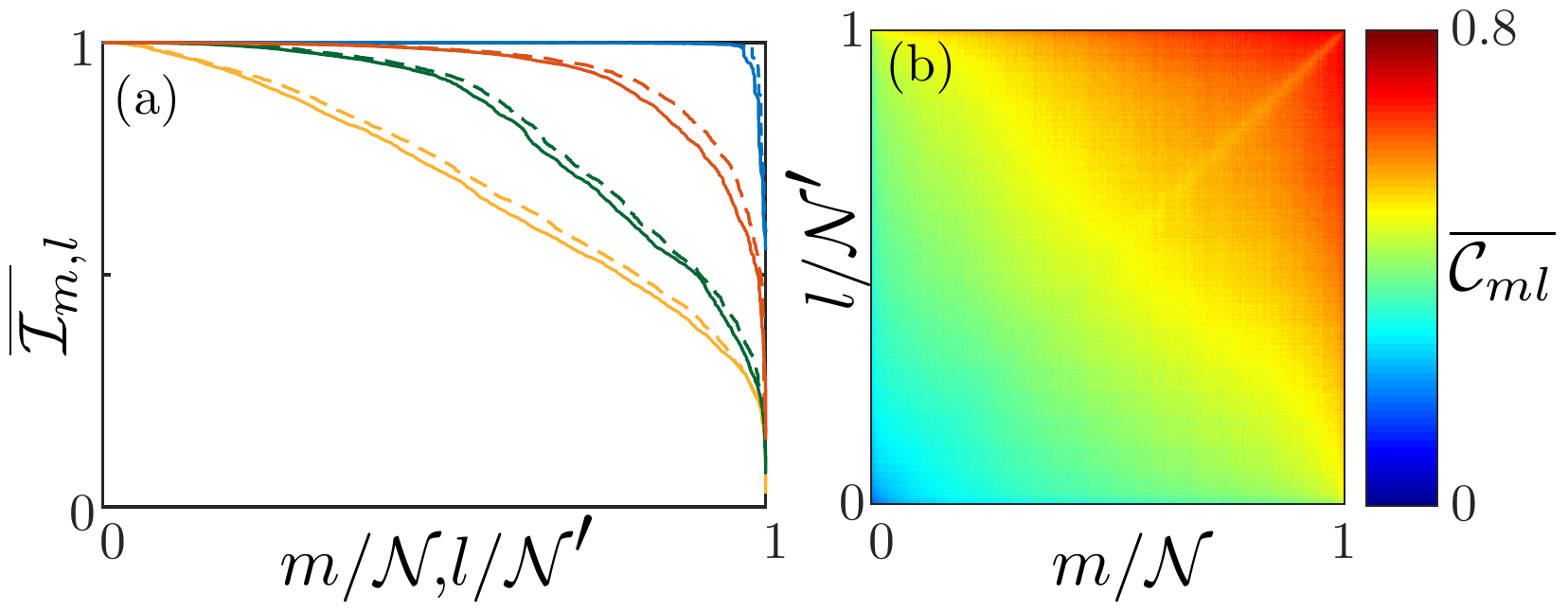}
	\caption{(a) Overlap $\mathcal{I}_{m,l}$ in individual realizations and (b) averaged orbital incompatibility in the system of Fig.~\ref{fig:OCM_L14_N7}, but with states taken from
half-filling (solid curves, $N=7$, index $m$) and from the sector with a particle removed (dashed curves, $N=6$, index $l$; total number of states denoted as $\mathcal{N}'$). See the text for an interpretation of the systematic correlations in (a) and reduced incompatibility along the diagonal of (b) in terms of emergent integrability.}
	\label{fig:OCM_L14_N7_N6}
\end{figure}

As shown in Fig.~\ref{fig:OCM_L14_N7_N6}, this feature is indeed present in the numerical data, and manifests itself both on the level of individual realizations as well as in the statistical averages. Firstly, as shown in panel (a),
in individual realizations, hence, for fixed given systems, the overlaps $\mathcal{I}_m$ of states with different particle number follow each other closely.
Secondly, as shown in panel (b), for pairs of states between which the generic compatibility is not already high, this interrelation shows up as a reduced averaged incompatibility of states along the diagonal.
The threshold value $\overline{\mathcal{C}_{ml}}\approx 0.5$ at the boundary of the region where this effect is visible reveals the residual incompatibility of the described pairs of states.

These geometric observations give quantitative insights into emergent integrability down to the level of individual systems, circumventing the challenging task of directly constructing the underlying quasi-particle operators.

\subsection{Analytic evidence based on a mininal case}
\label{sec:min}


To explain how the quasiparticle dressing arises, we provide a complete analytical solution of a minimal model, representing on a small system of $L=4$ sites with a particular disorder configuration $\varepsilon_1=\varepsilon_3$, $\varepsilon_2=\varepsilon_4$
(approximate disorder configurations of this type will be present generically in sufficiently long systems; however, for the analytical treatment we apply periodic boundary conditions). Shifting energies such that $\varepsilon\equiv\varepsilon_1=\varepsilon_3=-\varepsilon_2=-\varepsilon_4$, we can rewrite the Hamiltonian as
\begin{align}
	H=&-(d_1^\dagger d_2+d_2^\dagger d_1)
	+\varepsilon[n_1'-n_2'+n_3'-n_4' ]
	\nonumber\\
	&+V[(n_1'+n_3')(n_2'+n_4')-\sum_i n_i'+1],
\end{align}
where we changed the single-particle basis to
\begin{align}
	d_1^\dagger=\frac{c_1^\dagger+c_3^\dagger}{\sqrt{2}},\,\,\, d_2^\dagger=\frac{c_2^\dagger+c_4^\dagger}{\sqrt{2}},\,\,\,
	d_3^\dagger=\frac{c_1^\dagger-c_3^\dagger}{\sqrt{2}},\,\,\, d_4^\dagger=\frac{c_2^\dagger-c_4^\dagger}{\sqrt{2}}
	\label{eqn:dops}
\end{align}
and denoted the corresponding densities as $n_i'=d_i^\dagger d_i$.

In the interacting subspace of $N=2$ particles,
the eigenstates and associated energies are of the form
\begin{align}
	&|1\rangle=d_1^\dagger d_2^\dagger |\mathrm{vac}\rangle, \quad
	|2\rangle=d_3^\dagger d_4^\dagger |\mathrm{vac}\rangle  \quad(E_{1,2}=0),
	\nonumber\\
	&|3\rangle=\frac{1}{\sqrt{1+\alpha^2}}\,(\alpha d_1^\dagger- d_2^\dagger) d_3^\dagger |\mathrm{vac}\rangle \quad (E_3=\alpha),
	\nonumber\\
	&|4\rangle=\frac{1}{\sqrt{1+\alpha^2}}\,( d_1^\dagger+\alpha d_2^\dagger) d_3^\dagger |\mathrm{vac}\rangle \quad (E_4=-1/\alpha),
	\nonumber\\
	&|5\rangle=\frac{1}{\sqrt{1+\beta^2}}\, (\beta d_1^\dagger- d_2^\dagger) d_4^\dagger |\mathrm{vac}\rangle\quad (E_5=1/\beta),
	\nonumber\\
	&|6\rangle=\frac{1}{\sqrt{1+\beta^2}}\,(d_1^\dagger+ \beta d_2^\dagger) d_4^\dagger |\mathrm{vac}\rangle\quad (E_6=-\beta),
	\label{eqn:states}
\end{align}
where $|\mathrm{vac}\rangle$ denotes the state without any particles and
	\begin{align}
		\alpha=(\varepsilon-V/2)+\sqrt{1+(\varepsilon-V/2)^2},
		\nonumber\\
		\beta=(\varepsilon+V/2)+\sqrt{1+(\varepsilon+V/2)^2}.
\label{eqn:ab}
	\end{align}

Notably, these states are all exact Slater determinants,
but their form already suggests that they cannot be created from a common set of single-particle orbitals. To verify this feature, we identify the subspace of occupied orbitals in terms of the projectors $P_m$. In the  single-particle basis \eqref{eqn:dops}, where
$(P_m)_{ij}=\langle m|d_{i}^{\dag}d_{j}|m\rangle$,
this gives
\begin{align}
	P_1&=\left(\begin{array}{cccc}
		1&0&0&0\\0&1&0&0\\0&0&0&0\\0&0&0&0\\
	\end{array}\right),
	\quad
	P_2=\left(\begin{array}{cccc}
		0&0&0&0\\0&0&0&0\\0&0&1&0\\0&0&0&1\\
	\end{array}\right)
	\nonumber\\
	P_3&=\left(\begin{array}{cccc} \frac{\alpha^2}{1+\alpha^2}&-\frac{\alpha}{1+\alpha^2}&0&0\\-\frac{\alpha}{1+\alpha^2}&\frac{1}{1+\alpha^2}&0&0\\0&0&1&0\\0&0&0&0\\
	\end{array}\right)
	\nonumber\\
	P_4&=\left(\begin{array}{cccc}
		\frac{1}{1+\alpha^2}&\frac{\alpha}{1+\alpha^2}&0&0\\ \frac{\alpha}{1+\alpha^2}&\frac{\alpha^2}{1+\alpha^2}&0&0\\0&0&1&0\\0&0&0&0\\
	\end{array}\right)
	\nonumber\\
	P_5&=\left(\begin{array}{cccc} \frac{\beta^2}{1+\beta^2}&-\frac{\beta}{1+\beta^2}&0&0\\-\frac{\beta}{1+\beta^2}&\frac{1}{1+\beta^2}&0&0\\0&0&1&0\\0&0&0&0\\
	\end{array}\right)
	\nonumber\\
	P_6&=\left(\begin{array}{cccc}
		\frac{1}{1+\beta^2}&\frac{\beta}{1+\beta^2}&0&0\\ \frac{\beta}{1+\beta^2}&\frac{\beta^2}{1+\beta^2}&0&0\\0&0&1&0\\0&0&0&0\\
	\end{array}\right).
\end{align}

We now see that the projectors $P_3$ and $P_4$ do not commute with the projectors $P_5$ and $P_6$. The same is true in the original single-particle basis associated with the operators $c_i^\dagger$, for which the projectors follow from a unitary transformation $UP_lU^\dagger$ with
\begin{equation}
U=\frac{1}{\sqrt{2}}\left(\begin{array}{cccc} 1 & 0 & 1 & 0 \\ 0 & 1 & 0 & 1 \\ 1 & 0 & -1 & 0 \\ 0 & 1 & 0 & -1
\end{array}\right).
\end{equation}
In particular, any such single-particle basis change leaves the incompatibility measure \eqref{eqn:OPDM_5} invariant.

Therefore, this minimal model provides an exact realization of orbital incompatibility. Each projector uniquely identifies the corresponding Slater-determinant eigenstate up to an overall phase factor, but as the projectors do not commute, these Slater determinants cannot be built from a common set of single-particle orbitals \footnote{The states $|1\rangle$ and $|2\rangle$ are degenerate, but this cannot be used to remove this feature.}.


Notably, despite the described orbital incompatibility, we can still identify mutually commuting conserved quantities, beyond the number operator $I_0=n_1+n_2+n_3+n_4$, that provide a complete set of quantum numbers to uniquely discriminate all states. Within the two-particle sector, a possible choice is
\begin{align}
	I_1=&d_3^\dagger d_3,\quad
	I_2=d_4^\dagger d_4, \nonumber\\
	I_3=&\frac{1}{1+\alpha^2}d_3^\dagger d_3(\alpha d_1^\dagger- d_2^\dagger)(\alpha d_1- d_2)
	\nonumber\\
	&+\frac{1}{1+\beta^2}d_4^\dagger d_4(\beta d_1^\dagger- d_2^\dagger)(\beta d_1- d_2),
\end{align}
each of which have eigenvalues $0$ and $1$ and thus qualify as bit-like conserved quantities. We note that $I_3$ displays explicit dressings with occupations $d_i^\dagger d_i$.
As shown in App.~\ref{app:qparticles}, these considerations can be extended to explicitly formulate such bit-like conserved quantities that are valid in all particle-number sectors, which exhibit an analogous dressing with occupation operators of single-particle orbitals.

\section{Conclusions}
In summary, we established a geometric framework to describe the interrelation of eigenstates in fermionic many-body localized systems,  unraveling the intriguing structure and correlations exhibited by these paradigms of constrained complex quantum dynamics.
Applied to a paradigmatic model system, this approach reveals that while individual eigenstates are well approximated by single Slater determinants, they collectively depart from a uniquely-defined single-particle picture. This supports the notion of strongly dressed quasiparticle excitations residing behind the local conserved quantities that characterize the emergent integrability of these systems. The results complement the existing phenomenology based on individual eigenstates, such as the emergence of an area law of entanglement, both on the level of individual realisations as well as in statistical averages.

These considerations can also be usefully extended in a number of directions. For instance, based on our approach, one could inquire how well a single basis of orbitals can approximate a larger number of eigenstates, if not all of them. For this, the geometric perspective could be deepened by exploiting the general mathematical properties of the projectors $P_m$, hence, the structure of the Grassmannian  $G(H_1,N)$ defined by the $N$ dimensional subspaces in the single-particle Hilbert space $H_1$.
Furthermore,
to obtain additional insights into the ergodic phase and transition region, one could make use of the fact that
the notion of Brueckner orbital could be extended to systematically approximate states by multiple Slater determinants.
Finally, the described connection of our approach to the ubiquitous notion of l-bits,
as well as generalizations of the one-particle density matrix \cite{PhysRevA.101.063617,PhysRevResearch.2.023118},
could be explored to extend these considerations to non-fermionic systems, such as bosonic systems and spin chains.

\begin{acknowledgments}
This research was funded by UK Engineering and
Physical Sciences Research Council (EPSRC) via Grant
Nos. EP/P010180/1 and EP/L01548X/1. Computer time was provided by
Lancaster University's High-End Computing facility.
\end{acknowledgments}

\appendix

\section{Appendix: Random-matrix benchmark of orbital incompatibility}
\label{app:rmt}
To determine how sizeable the numerically observed orbital incompatibility of MBL states is, we compare it to the extreme case of two random  Slater determinants $|S_m\rangle$ and $|S_l\rangle$, with mutually uncorrelated projectors $P_m$ and $P_l$, admitting for additional generality that they may represent states of possibly different particle number $N$ and $N'$. As before, we denote the dimensionality of the single-particle Hilbert space as $L$.

We set out to obtain a compact result for the averaged squared incompatibility,
\begin{align}
	\overline{\mathcal{C}_{ml}^2}&=2 \,\mathrm{tr}\,\overline{P_mP_l}-2 \,\mathrm{tr}\,\overline{P_mP_lP_mP_l}
	\\&=2 \,\mathrm{tr}\,\overline{(P_lP_mP_l)}-2 \,\mathrm{tr}\,\overline{(P_lP_mP_l)^2},
	\label{C2start}
\end{align}
where the expression in the second line exploits that any projector fulfills  $P_l^2=P_l$.
This is useful, as
for any pair of states, we can adopt a single-particle basis in which $P_l$ is diagonal and projects onto the first $N$ states of the basis.
In this basis, the combination
\begin{equation}
P_lP_mP_l=\left(\begin{array}{cc}X & 0 \\ 0 & 0 \end{array}\right)
\end{equation}
appearing  in Eq.~\eqref{C2start} has a finite subblock $X$ of size $N\times N$, and in terms of this
\begin{equation}
\overline{\mathcal{C}_{ml}^2}=2 \,\mathrm{tr}\,\overline{X}-2 \,\mathrm{tr}\,\overline{X^2},
\end{equation}
where the trace is now of matrices of size $N$.

Diagonalizing $P_m$ in this basis, we can further write $(P_m)_{rs}=\sum_{n=1}^{N'} U_{rn}U^{*}_{sn}$, where the columns of the unitary matrix $U$ are the orthonormalized eigenstates of $P_m$, and the sum runs over the $N'$ eigenstates with eigenvalue 1 (the remaining eigenvalues vanish). Therefore, $X=uu^\dagger$ be further written in terms of a rectangular
$N\times N'$-dimensional subblock $u$ of the $L\times L$-dimensional unitary matrix $U$. We then have to  calculate
\begin{align}
	\overline{\mathcal{C}_{ml}^2}=2 \,\mathrm{tr}\,\overline{uu^\dagger}-2 \,\mathrm{tr}\,\overline{(uu^\dagger)^2}.
	\label{C2next}
\end{align}

In random-matrix theory, we can evaluate these averages using a standard ensemble.
Here we consider two cases, systems with real orbitals and systems with complex orbitals.
For the case of real orbitals, we take $U$ to be uniformly distributed over the orthogonal group $O(L)$, which is also known as the circular real ensemble (CRE; distinct from the circular orthogonal ensemble in which $U=U^T$ is still complex) \cite{RevModPhys.87.1037}. For the
case of complex orbitals, we take $U$ to be uniformly distributed over the unitary group $U(L)$, which corresponds to the standard circular unitary ensemble (CUE).
In both ensembles, all matrix elements of $U$ are equivalent,
which allows us to decompose the averaged traces
\begin{align}
	\mathrm{tr}\,\overline{uu^\dagger}&=NN'A,\\
	\mathrm{tr}\,\overline{(uu^\dagger)^2}&=NN'[B+(N+N'-2)C+(N-1)(N'-1)D]
\end{align}
of the truncated matrices into a small number of fundamental terms,
\begin{align}
	A&\equiv\overline{|u_{rs}|^2},
	\quad
	B\equiv\overline{|u_{rs}|^4},
	\\
	C&\equiv\overline{|u_{rs}|^2|u_{rt}|^2}=\overline{|u_{sr}|^2|u_{tr}|^2} \quad (s\neq t),
	\\
	D&\equiv\overline{u_{rs}u^*_{ps}u_{pt}u^*_{rt}}\quad(r\neq p,s\neq t),
\end{align}
where unspecified indices are unconstrained.
The same equivalence of matrix elements also implies the following sum rules,
\begin{align}
	\mathrm{tr}\, \overline{UU^\dagger}&=L^2A=L,
	\\
	\mathrm{tr}\, \overline{(UU^\dagger)^2}&=L^2B+2L^2(L-1)C+L^2(L-1)^2D=L,
	\\{}
	\overline{[(UU^\dagger)_{11}]^2}&=LB+L(L-1)C=1.
\end{align}

These expressions can all be combined to express the desired average as
\begin{align}
	\overline{\mathcal{C}_{ml}^2}|_{\mathrm{RMT}}&=-2DNN'(L-N)(L-N').
\end{align}

Averages such as the ones presented above also appear
in quantum transport, where $u$ would be interpreted as a block of a scattering matrix and the given combinations determine, for instance, universal conductance fluctuations and shot noise \cite{RevModPhys.69.731}. Furthermore, such averages also appear in the dynamics of quantum-chaotic systems, where $U$ represents a time-evolution operator \cite{Haake_1996}, while truncated versions appear in the description of leaky systems
\cite{PhysRevLett.93.154102} and in Floquet-descriptions of quantum transport \cite{PhysRevB.68.115313}.
Here, $A$ and $B$ can be worked out by interpreting them as moments of a coordinate from a random $L$-dimensional unit vector, parameterized in hyperspherical coordinates, thereby relating it to integrals of the form
\begin{equation}
I_{n,m}=\int_0^\pi\cos^n(\varphi)\sin^m(\varphi)d\varphi,
\end{equation}
while $C$ and $D$ then follow from the given sum rules.
In the CRE,
we then have
\begin{align}
	&A=\frac{I_{2,L-2}}{I_{0,L-2}}=\frac{1}{L},\\
	&B=\frac{I_{4,L-2}}{I_{0,L-2}}=\frac{3}{L(L+2)},\\
	&C=\frac{1}{L(L+2)}, \\
	&D=-\frac{1}{(L-1)L(L+2)},
\end{align}
while in the CUE we have
\begin{align}
	&A=2\frac{I_{2,2L-2}}{I_{0,2L-2}}=
	\frac{1}{L},\\
	&B=2\frac{I_{4,2L-2}}{I_{0,2L-2}}+2\frac{I_{2,2L}I_{2,2L-3}}{I_{0,2L-2}I_{0,2L-3}}=\frac{2}{L(L+1)},\\
	&C=\frac{1}{L(L+1)}, \\
	&D=-\frac{1}{L(L^2-1)}.
\end{align}
(In both cases, $A$ also follows from the given sum rule, and they can be related to each other using $B|_{\mathrm{CUE}(L)}=
2[B|_{\mathrm{CRE}(2L)}+C|_{\mathrm{CRE}(2L)}]$.)

Therefore, as our final results, we obtain
\begin{align}
	\overline{\mathcal{C}_{ml}^2}|_{\mathrm{RMT}}
	=  \frac{2NN'(L-N)(L-N')}{(L-1)L(L+2)} \quad \mbox{(CRE)}
\end{align}
for the averaged squared incompatibility of random Slater determinants with real orbitals,
and
\begin{align}
	\overline{\mathcal{C}_{ml}^2}|_{\mathrm{RMT}}
	=
	\frac{2NN'(L-N)(L-N')}{L(L^2-1)} \quad \mbox{(CUE)}
\end{align}
for the case of complex orbitals.
For large systems near half filling, both expressions approach $\overline{\mathcal{C}_{ml}^2}|_{\mathrm{RMT}} \sim L/8\equiv C_0^2$.
Level repulsion between the eigenvalues of $X$ then guarantees that fluctuations about this value are suppressed \cite{RevModPhys.69.731}. Therefore, $C_0$ not only characterizes the squared incompatibility, but can be used as a benchmark for the incompatibility $\mathcal{C}_{ml}$ itself.

\section{Explicit form of the quasiparticle operators in the minimal model}
\label{app:qparticles}
We here identify the exact form of quasiparticles for the minimal model of Sec.~\ref{sec:min} across all particle-number sectors.  These can be obtained by a unitary transformation \begin{equation}
q^\dagger_i = Ud_i^\dagger U^\dagger,
\label{eq:qparticletrafo}
\end{equation}
constrained by the requirement that the eigenstates in all particle-number sectors (including the vacuum state) can be written as
\begin{equation}
|n\rangle=\prod_{i\in I_n}q^\dagger_i | \mathrm{vac}\rangle,
\label{eq:qparticleproductstate}
\end{equation}
where $\{q_i^\dagger,q_j\}=\delta_{ij}$.

Using the exact form \eqref{eqn:states} of the eigenstates and eigenenergies, this unitary transformation can be identified as
	\begin{align}
	U
	=&n_1'n_2'+\bar n_1'\bar n_2'\nonumber\\
	&+ (D_1^\dagger d_1 \bar n_2'+D_2^\dagger d_2\bar n_1')\bar n_3'\bar n_4' \nonumber \\
	&
	+(D_1 d_1^\dagger n_2'+D_2 d_2^\dagger n_1')  n_3' n_4'
\nonumber \\
	&+  (D_{13}^\dagger d_1 \bar n_2'+ D_{23}^\dagger d_2 \bar n_1')n_3'\bar n_4'   \nonumber \\
	&
	+(D_{14}^\dagger d_1 \bar n_2'+ D_{24}^\dagger d_2 \bar n_1')\bar n_3'n_4',
	\end{align}
which features occupations $ n_i'=1-\bar n_i'=d_i^\dagger d_i$, as well as additional single-particle operators
\begin{align}
D_1^\dagger&=\frac{1}{\sqrt{1+\gamma^2}}(\gamma d_1^\dagger- d_2^\dagger),\\
D_2^\dagger&=\frac{1}{\sqrt{1+\gamma^2}}(d_1^\dagger+ \gamma d_2^\dagger)
\end{align}
with $\gamma=\varepsilon+\sqrt{1+\varepsilon^2}$,
and
\begin{align}
D_{13}^\dagger&=\frac{1}{\sqrt{1+\alpha^2}}(\alpha d_1^\dagger- d_2^\dagger),\\
D_{23}^\dagger&=\frac{1}{\sqrt{1+\alpha^2}}( d_1^\dagger+ \alpha d_2^\dagger),\\
D_{14}^\dagger&=\frac{1}{\sqrt{1+\beta^2}}(\beta d_1^\dagger- d_2^\dagger),\\
D_{24}^\dagger&=\frac{1}{\sqrt{1+\beta^2}}( d_1^\dagger+ \beta d_2^\dagger)
\end{align}
with $\alpha$ and $\beta$ as given in Eq.~\eqref{eqn:ab}.

Utilizing also the corresponding occupations, $N_i=1-\bar N_i =D_i^\dagger D_i$,
the quasiparticle operators \eqref{eq:qparticletrafo} follow as
	\begin{align}
	q_1^\dagger &=
	(n_3' n_4' + \bar n_3'\bar n_4')D_1^\dagger +  n_3'\bar  n_4' D_{13}^\dagger +
	\bar n_3'n_4' D_{14}^\dagger;
	\\
	q_2^\dagger &=
	(n_3' n_4' + \bar n_3'\bar  n_4')D_2^\dagger  + n_3'\bar  n_4'  D_{23}^\dagger+
	\bar n_3'n_4' D_{24}^\dagger
	\\
	q_3^\dagger &=
	[ n_1' n_2' +\bar n_1'\bar n_2'  + (D_{23}^\dagger D_2 \bar N_1 +
	D_{13}^\dagger D_1 \bar N_2)  \bar n_4'
\nonumber\\  &
+ (\bar N_2 D_1^\dagger D_{14} +
	\bar N_1 D_2^\dagger D_{24})  n_4']d_3^\dagger;
	\\
	q_4^\dagger &=
	[n_1' n_2' +\bar n_1'\bar n_2'  + (D_{24}^\dagger D_2 \bar N_1 +
	D_{14}^\dagger D_1 \bar N_2) \bar n_3'
\nonumber\\  &+ (\bar N_2 D_1^\dagger D_{13} +
	\bar N_1 D_2^\dagger D_{23})  n_3']d_4^\dagger.
	\end{align}

	From these expressions, the densities $I_i=q_i^\dagger q_i$ then deliver the desired bit-like conserved quantities.
	The two densities
	\begin{align}
	I_1=q_3^\dagger q_3=d_3^\dagger d_3,\quad I_2=q_4^\dagger q_4=d_4^\dagger d_4
	\end{align}
	take the simple form of a one-particle density. We furthermore have $I_3+I_4=d_1^\dagger d_1+d_2^\dagger d_2$, so that $I_1+I_2+I_3+I_4=I_0$ is the number operator.
	The remaining combination reads
	\begin{align}
	I_3-I_4 = &(N_1 - N_2)  (n_3'n_4'+ \bar n_3'\bar  n_4') + (N_{13} - N_{23}) n_3'\bar n_4'
\nonumber\\  &
	+ (N_{14} - N_{24}) \bar n_3' n_4'.
	\end{align}
As anticipated,  this is manifestly many-body-like, and displays a strong dressing with the particle-number occupations in various orbitals of the system.


%

\end{document}